\documentclass[12pt, a4paper]{article}

\usepackage{amsmath, amssymb, mathrsfs, newtxtext, newtxmath, tikz, enumitem, ulem}
\usepackage[margin=1in]{geometry}
\usepackage[colorlinks, urlcolor=black, citecolor=blue, link color=black]{hyperref}

\allowdisplaybreaks

\begin{document}

\title{ {\bf Comments on ``On the Dirac-Majorana neutrinos distinction in four-body decays''} (arXiv:2305.14140,
 Phys. Rev. D \textbf{109}, no.3, 033005 (2024))}

\author{C.~S.~Kim$^{1}$\thanks{cskim@yonsei.ac.kr}~, M.~V.~N.~Murthy$^{2}$\thanks{murthy@imsc.res.in}~~and Dibyakrupa~Sahoo$^{3}$\thanks{Dibyakrupa.Sahoo@fuw.edu.pl}}

\date{{\normalsize \it $^1$~Department of Physics and IPAP, Yonsei University, Seoul 03722, Korea\\$^2$~The Institute of Mathematical Sciences, Taramani, Chennai 600113, India\\$^3$Faculty of Physics, University of Warsaw, Pasteura 5, 02-093 Warsaw, Poland}\\~\\ \textrm{\today}}

\maketitle

\begin{abstract}
In Ref.~\cite{Marquez:2023rpc} the authors analyze the radiative
leptonic decay $\ell^- \to \nu_\ell \, \overline{\nu}_{\ell'} \,
\ell^{\prime -} \, \gamma$ to distinguish between Dirac and Majorana
nature of neutrinos. They utilize the back-to-back kinematics for this
purpose, a special kinematic configuration which we first proposed in
our paper \cite{Kim:2021dyj}. Here we point out how and why their
analysis of the back-to-back configuration is incorrect. This makes
their conclusion and comments invalid and untenable.
\end{abstract}

\section{Back-to-back kinematics}

Consider the 4-body decay of Ref.~\cite{Marquez:2023rpc} $\ell^- (p_1)
\to \nu_\ell (p_2) \, \overline{\nu}_{\ell'} (p_3) \, \ell^{\prime -}
(p_4) \, \gamma (p_5)$ or the 4-body decay of Ref.~\cite{Kim:2021dyj}
$B^0 \to \mu^- (p_-) \, \mu^+ (p_+) \, \overline{\nu}_\mu (p_1) \,
\nu_\mu (p_2)$, where the parent particle is either unpolarized or has
spin 0. These decays have a few common features.
\begin{enumerate}
\item In both these cases, the full phase space of the 4-body decay
can be described by using only $5$ \textit{independent variables}.%
\item The full angular distribution of the final particles can be
described by $3$ angles, which can be conveniently defined by
constructing two planes taking the 3-momenta of the final particles
into account.%
\item In the rest frame of the parent particle, if two final particles
fly away with 3-momenta of equal magnitudes but opposite directions
(i.e.\ back-to-back), then the other two final particles also fly away
back-to-back due to conservation of 3-momentum. This kinematic
configuration is referred to as the back-to-back (or b2b)
configuration.
\end{enumerate}

Ref.~\cite{Marquez:2023rpc} claims that our paper \cite{Kim:2021dyj}
made a mistake by taking the angle between the two planes to be zero
for b2b case, and they recommend that one should instead vary this
angle from $0$ to $2 \pi$ as an independent variable. Here we point
out how and why this is incorrect.

\subsection{From geometric point of view}

For convenience of discussion let us consider the decay $B^0 \to \mu^-
(p_-) \, \mu^+ (p_+) \, \overline{\nu}_\mu (p_1) \, \nu_\mu (p_2)$. In
the rest frame of the parent particle $B^0$, there are in general
three ways to define a pair of planes:
\begin{enumerate}[label=(\roman*)]
\item one plane formed by $\mu^+(\vec{p}_+), \mu^-(\vec{p}_-)$ and the
other plane formed by $\nu_\mu (\vec{p}_1), \overline{\nu}_\mu
(\vec{p}_2)$ with relative angle $\phi$, or%
\item one plane formed by $\mu^+(\vec{p}_+), \overline{\nu}_\mu
(\vec{p}_1)$ and the other plane formed by $\mu^- (\vec{p}_-), \nu_\mu
(\vec{p}_2)$ with relative angle $\phi'$, or%
\item one plane formed by $\mu^+ (\vec{p}_+), \nu_\mu (\vec{p}_2)$ and
the other plane formed by $\mu^- (\vec{p}_-), \overline{\nu}_\mu
(\vec{p}_1)$ with relative angle $\phi''$.
\end{enumerate}%
The above three ways of defining the pairs of planes are equivalent,
i.e.\ the final result should not depend on which set of planes one
chooses to work with. Let us consider the b2b condition $\vec{p}_+ +
\vec{p}_- = \vec{0} = \vec{p}_1 + \vec{p}_2$ in the rest frame of
$B^0$. This ensures that the directions of flights of the final muons
and neutrinos form two straight lines which intersect at the point
where $B^0$ would sit at rest before decay. Thus the final 3-momenta
in b2b case form only a single decay plane. Out of the above mentioned
three pairs of decay planes, if any pair have to correctly describe
this final physical plane in b2b case, then the relative angle between
the two planes must be zero, i.e.\ $\phi = \phi'= \phi'' = 0$. This is
an \textit{irrefutable} and \textit{unambiguous} fact. If $\phi \neq
0$ or $\phi' \neq 0$ or $\phi'' \neq 0$, there is no way to satisfy
the back-to-back conditions (in the local Euclidean geometry).
Therefore, one can not integrate out the angle between the planes and
must fix it to zero for b2b case, unlike what is done in
Ref.~\cite{Marquez:2023rpc}, where authors integrated the angle $\phi$
from 0 to $2 \pi$. As shown in our paper \cite{Kim:2021dyj} in
Eqs.~(33-35) of Sec.~IV~(F), if one integrates over $\phi$, the
difference between Dirac and Majorana cases would vanish.

\subsection{From counting number of independent variables}

As mentioned before, one only requires $5$ \textit{independent}
variables to describe the general 4-body kinematics of both
\cite{Marquez:2023rpc} and \cite{Kim:2021dyj}. However, the b2b
kinematics specified by $\vec{p}_+ + \vec{p}_- = \vec{0} = \vec{p}_1 +
\vec{p}_2$ has $3$ additional constraint equations, which implies that
only $2$ out of the initial $5$ variables would remain
\textit{independent} in b2b case, i.e.\ $3$ variables should either be
fixed by the b2b conditions or they must be dependent on the $2$
variables which are independent.

However, Eqs.\ (B7) and (B8) of Ref.~\cite{Marquez:2023rpc} have 4
\textit{independent} variables\footnote{The authors of
Ref.~\cite{Marquez:2023rpc} neglect mass of the final lepton $\ell'$
and use the conservation of energy to get $\beta_{\ell'} = E_{\ell'}
=\frac{1}{2} m_\ell - E_\nu$ which gets rid of one independent
variable.} $\theta_\nu$, $\theta_{\ell'}$, $\phi$ and $\beta_\nu =
E_\nu$. Thus in (B7) and (B8) one has too many arbitrary degrees of
freedom to describe the b2b case, against the fact that one requires
only $2$ \textit{independent} variables overall for this purpose. We
can see that not only any value of $\phi$, for which authors actually
claimed,   but also any value of $\theta_\nu$, $\theta_{\ell'}$ and
$E_\nu$ are also allowed for b2b case from these equations. 
Therefore, Eqs.~(B7) and (B8) are grossly misleading, and their claim
that ``... show cleverly that $\phi=0$ is not a constraint imposed by
the b2b kinematics ...'' is erroneous.

\subsection{From kinematics point of view}

The Eqs.~(B7) and (B8) of Ref.~\cite{Marquez:2023rpc} to start with
are unphysical as the coordinate axes have not been meaningfully
defined first for b2b case in Ref.~\cite{Marquez:2023rpc}. The
$z$-axis which is well defined in the general kinematics, $\hat{z} =
\left(\Vec{p}_1 + \Vec{p}_2\right)/\left| \Vec{p}_1 + \Vec{p}_2
\right|$ becomes undefined in b2b case as $\Vec{p}_1 + \Vec{p}_2 =
\Vec{0}$. Without the direction of $z$-axis fixed, the two decay
planes and the direction of $x$- and $y$-axes also remain undefined.
So before one can write the components of the 3-momenta along the
$x,y,z$-axes one must establish a physically meaningful connection
between the axes defined in the general kinematics and the axes one
would have in the b2b case. However Ref.~\cite{Marquez:2023rpc} did
not do this while writing down Eqs.~(B7) and (B8) from Eqs.~(B3) to
(B6). We do this precisely in Sec.IV.H of our paper
\cite{Kim:2021dyj}. During this fixing of the coordinate axes for b2b
case, we necessarily fix $3$ variables and are left with only $2$
independent variables, e.g.\ $E_\mu$ and $\theta$, as it should be.

One needs to first start by acknowledging the fact that if the two
planes defined in general kinematics were to correspond to the single
plane in b2b kinematics, then the angle $\phi$ between the two planes
must be set to zero.

\section{Conclusion}

Due to the wrong interpretation of back-to-back configuration by not
fixing the angle $\phi$ between the planes to be zero and by
integrating over $\phi$ from $0$ to $2\pi$, the statistical difference
between the Dirac and Majorana cases certainly vanishes in
Ref.~\cite{Marquez:2023rpc}. We have very clearly shown in our paper
\cite{Kim:2021dyj} in Eqs.~(33)--(35) of Sec.~IV~(F) that if one
integrates over $\phi$, the difference between Dirac and Majorana
cases would vanish. As far as experimental realization of the
back-to-back condition is concerned, we refer to the discussion in
Appendix A.1 of Ref.~\cite{Kim:2023iwz}.

More over, one is only left to wonder how fixing $\phi=0$ leads to
branching ratios, Eqs.~(C2) and (C3), which are roughly $6$ orders of
magnitude larger than the branching ratios when one integrates over
$\phi$ from $0$ to $2\pi$, Eq.~(C5) of Ref.~\cite{Marquez:2023rpc}.

In summary, we find the analysis of back-to-back kinematics by
Ref.~\cite{Marquez:2023rpc} as inaccurate and misleading. We disagree
with the incorrect conclusions of Ref.~\cite{Marquez:2023rpc} that one
should not fix $\phi$ to zero but integrate from $0$ to $2\pi$.
Therefore, we conclude by emphasizing that we clearly pointed out how
and why the analysis of the back-to-back configuration in
Ref.~\cite{Marquez:2023rpc} is incorrect, which makes their claims and
comments on our papers \cite{Kim:2021dyj, Kim:2023iwz} invalid and
untenable.

\section*{Acknowldgements}

The work of CSK is supported by NRF of Korea (NRF-2022R1A5A1030700
and NRF-2022R1I1A1A01055643). The work of DS is supported by the
Polish National Science Centre under the grant number
DEC-2019/35/B/ST2/02008.

\end{document}